# NONTRIVIAL MAGNETIC FIELD RELATED PHENOMENA IN THE SINGLE-LAYER GRAPHENE ON FERROELECTRIC SUBSTRATE

**(Author review)**

*Maksym V. Strikha[1*], Anna N. Morozovska[2†] and Zhanna G. Zemska[3]*

[1] *Taras Shevchenko Kyiv National University, Radiophysical Faculty*

*pr. Akademika Hlushkova 4g, 03022 Kyiv, Ukraine*

[2] *Institute of Physics, National Academy of Sciences of Ukraine,*

*pr. Nauky 46, 03028 Kyiv, Ukraine*

[3] *Taras Shevchenko Kyiv National University, Physical Faculty*

*Prospect Akademika Hlushkova 4a, 03022 Kyiv, Ukraine*

**Abstract**

The review is focused on our recent predictions of nontrivial physical phenomena taking place in the nanostructure single-layer grapheme on ferroelectric substrate, which are related with magnetic field. In particular we predicted that 180-degree domain walls in a strained ferroelectric film can induce p-n junctions in a graphene channel and lead to the unusual temperature and gate voltage dependences of the perpendicular modes $v_\perp$ of the integer quantum Hall effect. The non-integer numbers and their irregular sequence principally differ from the conventional sequence $v_\perp = 3/2, 5/3…$. The unusual $v_\perp$-numbers originate from significantly different numbers of the edge modes, $v_1$ and $v_2$, corresponding to different concentration of carriers in the left ($n_1$) and right ($n_2$) ferroelectric domains of p-n junction boundary. The difference between $n_1$ and $n_2$ disappears with the vanishing of the film spontaneous polarization in a paraelectric phase, which can be varied in a wide temperature range by an appropriate choice of misfit strain originated from the film-substrate lattice mismatch.

Next we studied the electric conductivity of the system ferromagnetic dielectric - graphene channel - ferroelectric substrate. The magnetic dielectric locally transforms the band spectrum of graphene by inducing an energy gap in it and making it spin-asymmetric with respect to the free electrons. It was demonstrated, that if the Fermi level in the graphene channel belongs to energy intervals, where the graphene band spectrum, modified by EuO, becomes sharply spin-asymmetric, such a device can be an ideal non-volatile spin filter. The practical application of the system under consideration would be restricted by a Curie temperature of a ferromagnet. Controlling of the Fermi level (e.g. by temperature that changes ferroelectric polarization) can convert a spin filter to a spin valve.

---

[*] e-mail: maksym.strikha@gmail.com
[†] e-mail: anna.n.morozovska@gmail.com

# НЕТРИВІАЛЬНІ ЯВИЩА В ОДНОШАРОВОМУ ГРАФЕНІ НА СЕГНЕТОЕЛЕТРИЧНІЙ ПІДКЛАДЦІ, ЯКІ ЗАЛЕЖАТЬ ВІД МАГНІТНОГО ПОЛЯ (АВТОРСЬКИЙ ОГЛЯД)

М.В. Стріха, Г.М. Морозовська, Ж.Г. Земська

**Резюме**

Огляд зосереджено на зроблених нами останнім часом передбаченнях незвичайних фізичних явищ у одношаровому графені на сегнетоелектричнній підкладці, що залежать від магнітного поля. Зокрема, ми передбачили, що 180-градусні доменні стінки в напруженій плівці сегнетоелектрика можуть індукувати p-n переходи у графеновому каналі й призвести до незвичайної залежності числа перпендикулярних мод $v_\perp$ у цілочисельному квантовому ефекті Холла від температури і напруги на затворі. Ці нецілочисельні числа та їхня порушена послідовність принципово відрізняються від звичайної послідовності $v_\perp = 3/2, 5/3…$. Незвичайні числа $v_\perp$ є наслідком суттєвої різниці між числами крайових мод $v_1$ і $v_2$, що відповідають різним концентраціям носіїв ліворуч ($n_1$) і праворуч ($n_2$) від p-n переходу, індукованого доменною стінкою. Різниця між $n_1$ і $n_2$ зникає разом зі зникненням спонтанної поляризації після переходу до параелектричної фази, що може відбуватися в широкому діапазоні температур, зумовленому відповідним вибором напруження невідповідності, що спричинене різницею сталих граток графенового каналу й сегнетоелектричної підкладки.

Надалі ми вивчали електропровідність системи феромагнітний діелектрик – графеновий канал – сегнетоелектрична підкладка. Феромагнітний діелектрик локально трансформує зонний спектр графену, індукуючи в ньому заборонену зону і роблячи його спін-асиметричним щодо вільних електронів. Ми показали, що, коли рівень Фермі в графеновому каналі потрапляє в енергетичні інтервали, у яких зонний спектр графену, модифікований EuO, робиться різко спін-асиметричним, такий пристрій може слугувати ідеальним енергонезалежним спіновим фільтром. Практичне застосування системи, яку ми розглядаємо, буде обмежене інтервалом температур, нижчих від температури Кюрі феромагнетика. Зміна енергії рівня Фермі (наприклад, під впливом температури, що змінює поляризацію сегнетоелектрика) може перетворити спіновий фільтр на спіновий клапан.

# 1. Introduction

After single layer graphene as a conducting channel for FET was obtained for the first time experimentally in 2004, many remarkable effects, caused by it's Dirac-like spectrum, like Klein paradox etc., were observed in deck-top experiments [1, 2, 3, 4], although before they were treated to be a part of high energy physics only. On the other hand, many other effects, like Quantum Hall Effect (QHE), which were observed previously in low temperature experiments only, were studied at ambient conditions; moreover, the unconventional integer QHE was predicted theoretically and observed experimentally [5, 6, 7]. In recent years special attention is paid to studies of graphene in different "smart" systems (like graphene on ferroelectric, see e.g. [8]) in regard of their possible application in ultrafast non-volatile electronic devices of new generation; and to graphene application in spintronics.

This our review is focused on recent predictions of unusual physical phenomena taking place in the nanostructure single-layer grapheme on ferroelectric substrate, which are related with magnetic field. In particular we predicted that 180-degree domain walls in a strained ferroelectric film can induce p-n junctions in a graphene channel and lead to the unusual temperature and gate voltage dependences of the perpendicular modes $v_\perp$ of the integer quantum Hall effect [9]. The non-integer numbers and their irregular sequence principally differ from the conventional sequence $v_\perp = 3/2, 5/3\ldots$. The unusual $v_\perp$-numbers originate from significantly different numbers of the edge modes, $v_1$ and $v_2$, corresponding to different concentration of carriers in the left ($n_1$) and right ($n_2$) ferroelectric domains of p-n junction boundary. The difference between $n_1$ and $n_2$ disappears with the vanishing of the film spontaneous polarization in a paraelectric phase, which can be varied in a wide temperature range by an appropriate choice of misfit strain originated from the film-substrate lattice mismatch.

Next we studied the electric conductivity of the system ferromagnetic dielectric - graphene channel - ferroelectric substrate [10]. The magnetic dielectric locally transforms the band spectrum of graphene by inducing an energy gap in it and making it spin-asymmetric with respect to the free electrons. It was demonstrated, that if the Fermi level in the graphene channel belongs to energy intervals, where the graphene band spectrum, modified by EuO, becomes sharply spin-asymmetric, such a device can be an ideal non-volatile spin filter. The practical application of the system under consideration would be restricted by a Curie temperature of a ferromagnet. Controlling of the Fermi level (e.g. by temperature that changes ferroelectric polarization) can convert a spin filter to a spin valve.

## 2. Integer quantum hall effect in a graphene channel with p-n junction at domain wall in a strained ferroelectric film

It had been demonstrated theoretically that the Dirac-like spectrum of graphene [1-4], and, consequently, additional double degeneration of a zero Landau level (**LL**), which is common for conduction and valence bands, results in a unconventional form of a integer quantum Hall effect (**IQHE**) [5-7]:

$$\sigma_{xy} = -\frac{e^2}{2\pi\eta}\nu, \qquad \nu = \pm 2(2k+1). \qquad (1.1)$$

Here $\sigma_{xy}$ is the xy-component of conductance tensor and $v$ is the number of edge modes-[6]. The Hall plateaus are centered around the values $\nu = \pm 2(2k+1)$, where $k = 0, 1, 2...$ is an integer. Nonzero k numbers are given by expression [6]:

$$k = \left[\frac{n}{4n_B} - \frac{1}{2}\right]. \qquad (1.2)$$

Symbol "[]" stands for the integer part of a number; $n$ is the 2D concentration of electrons in graphene channel, and $n_B = eB/(2\pi\eta)$ is the density of magnetic field flux, threading the 2D surface corresponding to the degree of the $k$-th LL occupation.

Abanin and Levitov [11] explained the peculiarities of IQHE observed in graphene with p-n junction across the conduction channel [12]. It has been demonstrated, that the electron and hole modes can mix at the p-n boundary in the bipolar regime, leading to current partition and quantized short noise. On the contrary, recently the formation of IQHE with p-n junctions created along the longitudinal direction of graphene cannel had been studied [13], and the enhanced conductance can be observed in the case of bipolar doping. In both cases (p-n junction along and across the channel) IQHE observation can be exploited to probe the behavior and interaction of quantum Hall channels.

Recently [14, 15] we have studied the conductivity of graphene channel with p-n junction induced by a 180-degree ferroelectric domain wall (**FDW**) in a ferroelectric substrate. P-N junction in graphene at FDW was studied experimentally in Refs. [16, 17]. Later on we have studied p-n junction dynamics induced in a graphene channel by a FDW motion in the substrate [18] and demonstrated the possibility how to vary a number of p-n junctions in a channel between the source and drain electrodes by the motion of FDW in ferroelectric substrate [19].

On the basement of these works we have studied [9] the peculiarities of IQHE in graphene channel with p-n junction at FDW in a strained ferroelectric film on a thick rigid substrate electrode. Noteworthy, that the temperature of phase transition from the ferroelectric (**FE**) to paraelectric (**PE**) phase taking place in a strained ferroelectric film can be changed by a misfit strain in a wide temperature range including room temperatures [20]. Utilizing the fact, we revealed the unusual features of the IQHE in graphene at room temperature, unknown earlier. Despite we have an integer character of QHE on both sides of FDW, the total number of perpendicular modes proves to be non-integer in a very unusual way. Let us focus on the result in more details.

System under consideration in Ref.[9] is shown in **Fig.1**. It includes top gate electrode, oxide dielectric layer, graphene conducting channel with source and drain electrodes, and a strained ferroelectric film epitaxially clamped to a thick rigid substrate electrode playing the role of the bottom gate. Misfit strain $u_m \approx \frac{a-b}{b}$, caused by the lattice constants mismatch between the ferroelectric film (with a lattice constant *a*) and its substrate (with a lattice constant *b*), exists at the film-substrate interface, and creates elastic strain in the film [21]. The magnetic field **B** is applied normally to the graphene channel plane. The spontaneous polarization changes its direction from $-P_S$ to $+P_S$ on different sides of the 180-degree FDW. The FDW induces a p-n junction in graphene at that the wall plane coincides with the p-n junction position. Here *d* is the thickness of oxide dielectric layer and *h* is the thickness of ferroelectric film.

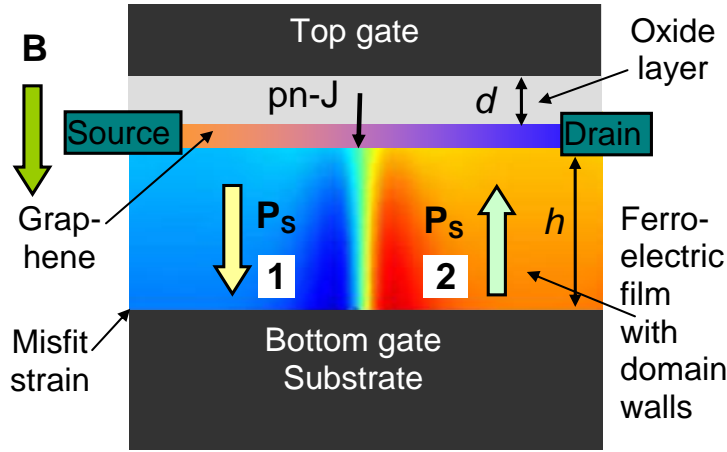

**Figure 1.** Schematics of the considered heterostructure "top gate/ dielectric oxide layer / graphene channel / strained ferroelectric film with a 180-degree domain wall / compressive (or tensile) substrate - bottom gate". The spontaneous polarization changes its direction from $-P_S$ to $+P_S$ on different sides of the FDW. The strong magnetic field **B** is applied normally to the graphene channel plane**.** (Reproduced from [M. V. Strikha, A. I. Kurchak, and A. N. Morozovska. Integer Quantum Hall Effect in Graphene Channel with p-n Junction at Domain Wall in Ferroelectric Substrate. J. Appl. Phys. **125**, 082525 (2019)], with the permission of AIP Publishing).

We regarded that the graphene channel is highly homogeneous (see e.g. Ref.[22] for experiment), and so the scattering at short-range graphene defects is much smaller than the scattering at surface ionized impurities. We consider a single-layer graphene at ambient conditions; therefore there should not be any significant thermal effect for graphene edge states, which are considered can within the approximation of sharp boundaries (see e.g. Refs.[23] and [11]). For a ferroelectric this is possible only if the thickness of 180-degree FDW is much smaller than the domain width and graphene channel length. To minimize the surface-induced domain wall broadening [24], the dead layer between the ferroelectric and graphene should be absent and the concentration of free carriers in graphene should be high enough to provide the effective screening of the spontaneous polarization in a ferroelectric. Since the modern interface engineering of epitaxial hetero-structures allows one to control the existence of dead layers and their properties [25], one can consider the temperature dependence of the 180-degree FDW intrinsic width $w(T)$ given by expression [25, 26] $w(T) = \dfrac{w_0}{\sqrt{1 - T/T_C^*}}$, where $w_0$ is the intrinsic width of uncharged domain wall at 0 K. Its value is about (0.3 – 0.5) nm for inorganic ferroelectrics such as BaTiO$_3$, Pb(Zr,Ti)O$_3$, LiNbO$_3$ and BiFeO$_3$.

The Curie temperature $T_C^*(u_m^*)$ of a ferroelectric strained film can be controlled by appropriate choice of the misfit strain $u_m$ between the film and substrate [20], and corresponding expression for the out-of-plane polarization has the form:

$$T_C^*(u_m^*) = T_C\left(1 + \dfrac{2Q_{12} u_m^*}{\alpha_T T_C (s_{11} + s_{12})}\right), \quad (1.3)$$

where $T_C$ is a Curie temperature of a bulk ferroelectric, $Q_{12}$ is the electrostriction coefficient, $s_{ij}$ are elastic compliances, positive coefficient $\alpha_T$ is expressed via inverse Curie-Weiss constant. Effective misfit strain $u_m^*$ is assumed to be equal to the film-substrate misfit strain, $u_m^* = u_m$, when the ferroelectric film is free from misfit dislocations or other elastic defects, which can be natural relaxation sources for the strains (e.g. the thickness $h$ is smaller than the critical thickness $h_d$ of misfit dislocations appearance) [21]. Under the presence of misfit dislocations at $h > h_d$ the approximation $u_m^*(h) \approx u_m \dfrac{h_d}{h}$ can be used [27]. The changes of the Curie temperature caused by misfit strain can reach several hundreds of Kelvins [20]. In accordance with expression (3) the transition temperature $T_C^*(u_m^*)$ can be made close to the room temperature by choosing appropriate misfit strain $u_m$ even for ferroelectrics with relatively high $T_C$ like PbZr$_x$,Ti$_{1-x}$O$_3$ (x<0.5) with $T_C$~(500 - 700 K), LiNbO$_3$ with $T_C$~1000 K and multiferroic BiFeO$_3$ with $T_C$~ 1100 K. To approach the room temperature for $T_C^*(u_m^*)$, the film should be thick enough for ferroelectrics with relatively low $T_C$, namely for Rochelle salt with $T_C$= 293 K, CuInP$_2$S$_6$ with $T_C$=(302 – 305)K, for PVDF with $T_C$= 438 K, and for BaTiO$_3$ with $T_C$= 381 K.

The concentration of carriers in a graphene channel placed at a single Ising-type 180-degree FDW on the left ($n_1(x)$) and right ($n_2(x)$) sides from the wall is defined by the difference of electric displacement normal components in the oxide layer and in the ferroelectric, $D_3^O(x, z = 0) - D_3^{FE}(x, z = 0)$, and so it acquires the form:

$$n_1(x) = \frac{1}{e}\left[D_3^O(x \leq 0, z = 0) - D_3^{FE}(x \leq 0, z = 0)\right], \quad x \leq 0, \quad (1.4a)$$

$$n_2(x) = \frac{1}{e}\left[D_3^O(x \geq 0, z = 0) - D_3^{FE}(x \geq 0, z = 0)\right]. \quad x \geq 0. \quad (1.4b)$$

Electric displacements are defined by the electric field $E_3 = -\partial\varphi/\partial z$ and ferroelectric polarization $P_3$ as $D_3^O = \varepsilon_0\varepsilon_O E_3^O$ and $D_3^{FE} = \varepsilon_0\varepsilon_{33}^b E_3^{FE} + P_3^{FE}$, where $\varepsilon_O$ is the relative dielectric permittivity of oxide layer, and $\varepsilon_{33}^b \sim (4 - 7)$ is the "background" permittivity of ferroelectric [28, 29]; $\varepsilon_0$ is dielectric permittivity of vacuum.

We derived [9] that approximate analytical expressions for the values $n_1$ and $n_2$,

$$n_{1,2}(T, u_m^*) \approx \frac{\varepsilon_0 V_g}{e}\left(\frac{\varepsilon_O}{d} + \frac{\varepsilon_{33}^f(T)}{h}\right) \pm \frac{P_S(T, u_m^*)}{e}, \quad (1.5)$$

are valid quantitatively with high accuracy (several %) far enough from the immediate vicinity of the FDW, where the spontaneous polarization profile continuously changes its value from $-P_S(T, u_m^*)$ to $+P_S(T, u_m^*)$. $V_g$ is top gate voltage, $\varepsilon_{33}^f(T)$ is its relative dielectric permittivity.

Corresponding number of edge modes, $v_{1,2}$, with carriers concentrations to the left and to the right sides of FDW boundary $n_{1,2}$, are

$$v_{1,2} = \pm 2(2k_{1,2} + 1), \qquad k_{1,2} = \left[\frac{n_{1,2}}{4n_B} - \frac{1}{2}\right], \quad (1.6)$$

Expressions (1.5)-(1.6) are valid until the spontaneous polarization exists, i.e. below the temperature $T_C^*$ of the ferroelectric-paraelectric phase transition, because FDW and induced p-n junction disappear at temperatures $T > T_C^*$.

Later we shall examine two cases: **(a)** p-n junction across the graphene conducting channel created by 180-degree domains; **(b)** p-n junction along the graphene conducting channel. For the cases the conductance in QHE regime is governed by a number of modes [11]

$$\nu_\perp = \frac{|\nu_1||\nu_2|}{|\nu_1|+|\nu_2|}, \qquad \nu_{\uparrow\uparrow} = |\nu_1|+|\nu_2|. \tag{1.7}$$

where $\nu_{1,2}$ are determined by Eqs.(1.5), (1.6) with different polarization directions ($\pm P_S$) corresponding to left and right sides of p-n junction.

The reasonable ranges of parameters in Eqs.(1.3)-(1.7) are the following: magnetic field strength $B = (0.5 - 10)$ Tesla, gate voltage $|V_g| = (0-10)$ V, oxide layer thickness $d = (5-50)$ nm and relative dielectric permittivity $\varepsilon_O = (1 - 100)$. The temperature dependence of the spontaneous polarization,

$$P_S(T, u_m^*, h) = P_0 \sqrt{1 - \frac{T}{T_{cr}(u_m, h)}}, \tag{1.8}$$

corresponds to the second order phase transition from the FE to PE phase. The FE-PE phase transition temperature $T_{cr}(u_m, h)$ of the strained ferroelectric film on a rigid substrate can be controlled by changing of misfit strain $u_m$ [20] and its thickness $h$ due to the finite size effects [32], namely, $T_{cr}(u_m, h) \approx T_C^*(u_m^*) - \frac{l_S}{\alpha_T \varepsilon_0 h}$, where $l_S$ is the screening length in graphene layer that is typically much smaller than 0.5 nm [30, 31]; and the expression for $T_C^*(u_m^*)$ is given by Eq.(1.3). $T_{cr}(u_m, h)$ can be made close to room temperature by using appropriate film-substrate pair, which defines the misfit strain $u_m$ and decreasing the film thickness below 100 nm due to the existence of the thickness-induced phase transition into a PE phase [32, 33]. The value $P_0 = \sqrt{\alpha_T T_C/\beta^*} \approx \sqrt{\alpha_T T_C/\beta}$ corresponds to the bulk ferroelectric, and, as a rule, it varies in the range $(0.1-1)$ C/m$^2$

Using above parameters one can estimate that the number of perpendicular modes $\nu_\perp$ can vary from 1 in the immediate vicinity of $T_{cr}(u_m, h)$, to very high values (>$10^2$) far from the Curie temperature. Actually two sharp symmetric stepped minimums are clearly seen at the voltage dependences of the modes $|\nu_1|$, $|\nu_2|$ and $\nu_\perp$ shown in **Fig. 2a**. The voltage position and the distance between the minimums is about 1.5 V for the difference $\Delta T = T_{cr}(u_m, h) - T = 5$ K. In the vicinity of each minima either $\nu_\perp \cong |\nu_1|$ or $\nu_\perp \cong |\nu_2|$ (red curves almost coincide either with black one or with blue ones). However the filling numbers of $|\nu_1|$ or $|\nu_2|$ are integers (2, 4, 6, etc), the number of perpendicular modes $\nu_\perp(V_g)$ appears non-integer and its denominator differs from 2 or 3 with the increase of either $|\nu_1|$ or $|\nu_2|$, that corresponds to the increase of the gate voltage (see the difference between red, black or blue curves in **Fig.2a**). The mode $\nu_{\uparrow\uparrow}$ is weakly voltage dependent.

The dependences of $|v_1|$, $|v_2|$, $v_\perp$ and $v_{\uparrow\uparrow}$ on oxide layer thickness $d$ are shown in **Fig.2b** for a fixed gate voltage $V_g$. A pronounced stepped minimum is seen on the dependence $v_\perp(d)$. The thickness dependences of either $|v_1|$ or $|v_2|$ are very close to the dependence $v_\perp(d)$ near the minimum. The values of $|v_1|$ or $|v_2|$ are integers equal to 2, 4, 10, etc, but the values of perpendicular modes $v_\perp(d)$ appears non-integer and its denominator differs from 2 or 3 with the increase of either $|v_1|$ or $|v_2|$.

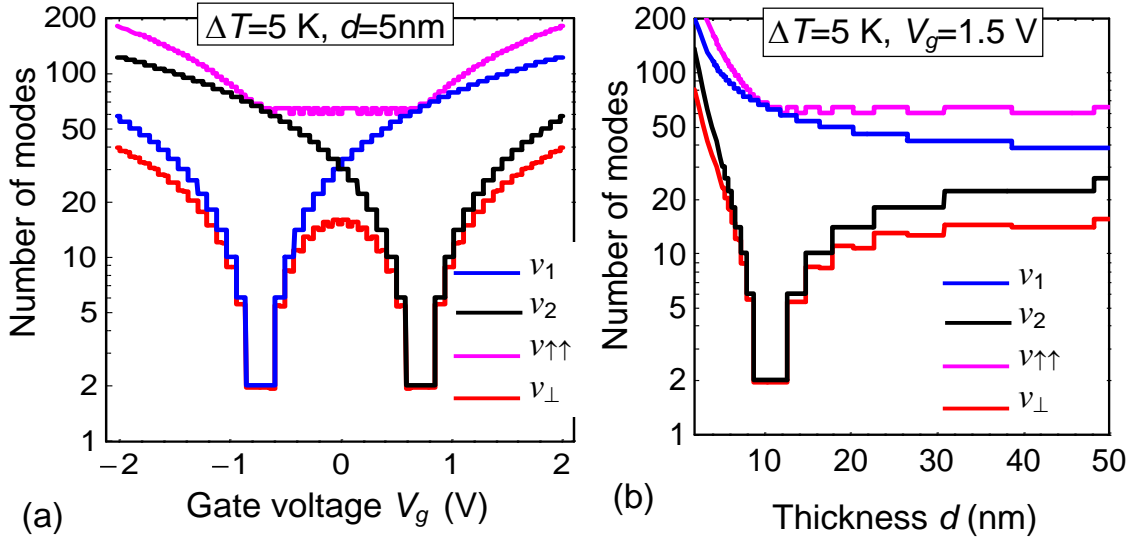

**Figure 2.** (a) The modes $|v_1|$ (blue curves), $|v_2|$ (black curves), $v_\perp$ (red curves) and $v_{\uparrow\uparrow}$ (magenta curves) in dependence on the gate voltage $V_g$ (in volts) calculated for $d = 5$ nm and the temperature difference $\Delta T = T_{cr}(u_m, h) - T$ equal to 5 K. (b) The modes $|v_1|$ (blue d curves), $|v_2|$ (black curves), $v_\perp$ (red curves) and $v_{\uparrow\uparrow}$ (magenta curves) in dependence on oxide layer thickness $d$ (in nm) calculated for $V_g = 1.5$V. Other parameters: $B$=10 Tesla, $\varepsilon_O = 10$, $T_{cr}(u_m, h)$=305 K and $P_0 = 0.1$ C/m². (Reproduced from [M. V. Strikha, A. I. Kurchak, and A. N. Morozovska. Integer Quantum Hall Effect in Graphene Channel with p-n Junction at Domain Wall in Ferroelectric Substrate. J. Appl. Phys. **125**, 082525 (2019)], with the permission of AIP Publishing).

From **Fig.2** the number of perpendicular modes $v_\perp$ varies from small integers to high non-integer numbers, in the vicinity of transition temperature from the ferroelectric to paraelectric phase. Thus, we predicted that 180-degree FDWs in a ferroelectric substrate, which induce p-n junctions in a graphene channel, lead to nontrivial temperature and gate voltage dependences of the perpendicular and parallel modes of the unconventional IQHE. Unexpectedly the number of perpendicular modes $v_\perp$, corresponding to the p-n junction across the graphene conducting channel, varies from integers to different non-integer numbers depending on the gate voltage, temperature and oxide layer thickness, e.g. $v_\perp$=1.94, 2,…5.1, 6.9, …9.1…, 23, 37.4 for the first Hall plateaus, at that smaller numbers corresponds to temperatures in the vicinity of transition from the ferroelectric to paraelectric phase.

### 3. Magnetic dielectric - graphene - ferroelectric system as a promising non-volatile device for modern spintronics

A field effect transistor (FET) with a graphene channel on a dielectric substrate was created in 2004 for the first time [34], and multiple attempts have been made to use the unique properties of the new 2D-material in spintronics [35, 36]. Shortly after it was concluded that the graphene is poorly attractive for spintronics, since the magnetoresistance of the graphene-based spin valve, $MR = \frac{R_{AP} - R_P}{R_P}$, where $R_{AP}$ and $R_P$ are the valve resistances for antiparallel and parallel orientation of magnetization at ferromagnetic contacts, are very small [35]. However, effective spin valves with a graphene "spacer" and cobalt contacts have been created soon [37], and intensive efforts have been made to improve them [38, 39], including the solution to use a single-layer graphene channel in an active ferromagnetic element [40]. To realize this, a dielectric ferromagnet EuO was imposed on the part of the graphene channel to induce the strong spin polarization of graphene π-orbitals. The splitting of the graphene band states into the subbands with the orientation of the spin values "up" and "down" occurs, and EuO induces the energy gap between these bands [41].

Recently [42] we have shown that it is possible to create a non-volatile spin valve / filter similar to that proposed in Ref.[40], where, however, the appropriate location of the Fermi level is provided not by the gate voltage, but by the spontaneous polarization of the ferroelectric substrate (see **Fig. 3a**). The single-layer graphene channel is considered as an infinitely thin two-dimensional (2D) gapless semiconductor of rectangular shape with length $L$ and width $W$. Since we regard that $L$ is smaller than the free path of the electron $\lambda$, the channel conductivity takes place in the ballistic regime.

The graphene channel is placed on a single-domain ferroelectric film with a spontaneous polarization $P_S$. Since the ferroelectric substrate can be used for the doping of a graphene conductive channel by a significant number of carriers without the traditional application of the gate voltage [43, 44, 45, 26, 47], $P_S$ value determines the specific carrier concentration in the channel, $n = P_S/e$, where $e$ is the electron charge. The sign "+" of $P_S$ corresponds to a positive bound charge at the graphene-ferroelectric interface, and thus the graphene doping with electrons. The sign "-" corresponds to the negative bound charge at the interface and to the channel doping with holes. As in Ref.[40], a magnetic dielectric EuO with a length of $l \ll L$ is superimposed on the graphene channel and the top gate is placed above the magnetic dielectric.

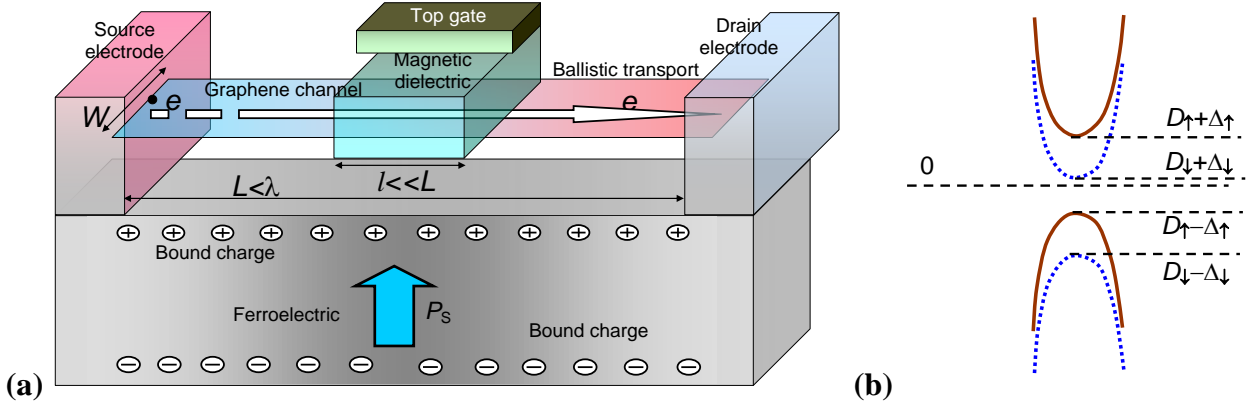

**Fig.3 (a)** Graphene single-layer of length $L$ and width $W$ is placed between ferromagnetic EuO and single-domain ferroelectric film. **(b)** The gap in the energy spectrum of graphene induced by a magnetic dielectric (adapted from Ref.[40]). Dotted curves correspond to the "down" spin states, solid curves are for states with spin "up". Zero energy "0" corresponds the position of the Dirac point in a nonmagnetic graphene channel. (Reproduced from [A. I. Kurchak, A. N. Morozovska, and M. V. Strikha. Magnetic dielectric- graphene- ferroelectric system as a promising non-volatile device for modern spintronics J. Appl. Phys. 125, 174105 (2019)], with the permission of AIP Publishing).

Isolated single-layer graphene is a 2D gapless semiconductor with the linear band spectrum near the Dirac point [35], $E_\pm(k) = \pm \eta v_F k$, where $k = \sqrt{k_x^2 + k_y^2}$ is the wave vector value, $v_F = 10^6$ m/s is a Fermi velocity, and the signs "+" and "-" correspond to the conduction and valence bands, respectively. In the graphene channel section located under EuO the spectra [41] becomes:

$$E_\sigma(k) = D_\sigma \pm \sqrt{(\eta v_\sigma k)^2 + (\Delta_\sigma / 2)^2}, \qquad (2.1)$$

where the subscript $\sigma$ designates the two values of the spin projection ($\sigma = \uparrow, \downarrow$); the energies of the Dirac point shift are $D_\uparrow = 31$ meV and $D_\downarrow = -31$ meV, respectively. The energies determining the splitting are $\Delta_\uparrow = 134$ meV and $\Delta_\downarrow = 98$ meV; and the renormalized values of the electron velocity are equal to $v_\uparrow = 1.15 \cdot 10^6$ m/s and $v_\downarrow = 1.40 \cdot 10^6$ m/s, respectively [40]. The gap in the energy spectrum of graphene induced by a dielectric ferromagnet is shown in **Fig. 3b**.

Note here, that first principle calculations [40] demonstrate strong doping of graphene by electrons from EuO, leading to the 1.4 eV shift of Fermi level in graphene on EuO into conduction band. Therefore a top gate is imposed over the magnetic dielectric in order to compensate by a proper applied gate voltage a local doping of graphene channel by electrons from proximity effect of EuO. Moreover, we treat the case of the short graphene on EuO section $l \ll L$. Therefore the doping of graphene conducting channel along all it's length $L$ is determined by the ferroelectric spontaneous polarization. This means that the position of the Dirac point "0" in a nonmagnetic graphene channel is twice closer to the edge of the upper band than to the lower one [see Eq.(2.1)], and so there is a certain asymmetry in the spin-polarized spectra under EuO for the ferroelectric polarization up and down directions.

Although the electron would pass through a non-magnetic graphene channel of length $L$ without dispersion, however, the presence of the section with length $l \ll L$ in the channel, where the graphene has pronounced magnetic properties, leads to the necessity to account for the local scattering of carriers with different spin signs. The conductivity of the graphene channel, taking into account the double degeneration of graphene at points $K, K'$, will be described by the modified Landauer formula [46, 47]:

$$G = \sum_\sigma G_\sigma, \qquad G_\sigma = \frac{2e^2}{\pi\eta} M(E_F) T_\sigma(E_F), \qquad (2.2)$$

where $M(E_F)$ is the number of conductance modes. The transmission coefficient $T_\sigma(E_F)$ is the probability that the electron will pass without a scattering the "magnetic" section of length $l$, which depends on the value of the electron spin. For the full conductivity it is necessary to sum for both spin values.

Using the relation $l \ll L$ we can assume that $M(E_F) = 0$ with high accuracy, when Fermi energy level is inside the energy gap of the spectrum (2.1), and lies between the 2-nd and 3-rd energies levels. Outside the gap $M(E_F)$ is described by the expression [47], $M(E_F) = Int\left[\frac{2W}{\lambda_{DB}}\right]$, and has the physical meaning of the number of de Broglie half-wavelengths $\lambda_{DB}/2$ for an electron in the graphene channel, which can be located at the width of this channel $W$. Symbol "Int" denotes the integer part.

Using relation between 2D-concentration of electrons and Fermi energy in graphene, $E_F = \eta v_F \sqrt{\pi n} = \eta v_F \sqrt{\frac{\pi P_S}{e}}$, we obtain for the de Broglie wavelength:

$$\lambda_{DB} = \frac{2\pi \eta v_F}{E_F} = 2\sqrt{\frac{\pi e}{P_S}}. \qquad (2.3)$$

From Eq.(2.3), the possible range of Fermi energy tuning by the spontaneous polarization varying in the range $P_S \cong (0.01 - 1)$ C/m$^2$ is $(0.3 - 3.0)$ eV (for $v_F \approx 10^6$ m/s and $\eta = 6.583 \times 10^{-16}$ eV·s).

When the Fermi level lies outside the gap of the spectrum (2.1), the number of conductance modes is:

$$M(E_F) = Int\left[W\sqrt{\frac{P_S}{\pi e}}\right] \qquad (2.4)$$

It was shown in Ref.[40] that the transmission coefficient $T_\uparrow(E_F) \approx 0$ and $T_\downarrow(E_F) \approx 1$ when the Fermi level is within the energy interval between the first and the second energy levels in Eq.(2.1). When the Fermi level falls into the interval between the third and the fourth energy values, $T_\uparrow(E_F) \approx 1$ and $T_\downarrow(E_F) \approx 0$. When the values of polarization are so high that the Fermi level lies either above the first or below the fourth energy levels, both transmission coefficients are close to unity.

Thus, there are intervals of ferroelectric polarization, for which the considered system is an ideal spin filter, and the value of the ratio $\frac{G_\uparrow - G_\downarrow}{G_\uparrow + G_\downarrow}$ is close to 1 for sufficiently small polarizations of the order of mC/m².

The value of the spontaneous polarization $P_S$ in thin ferroelectric films can be controlled by size effect, temperature $T$ and/or misfit strain $u_m$ originated from the film-substrate lattice constants mismatch [9, 42, 48, 49]:

$$P_S(T, u_m, h) = P_0 \sqrt{1 - \frac{T}{T_{cr}(u_m, h)}}, \qquad (2.5a)$$

The spontaneous polarization $P_0$ corresponds to the bulk sample at zero Kelvin; and can vary $P_S$ from the bulk values of (0.5 – 0.05) C/m² order to the values hundreds of times smaller. The temperature $T_{cr}(u_m, h)$ of the second order phase transition of the film to the paraelectric state has the form [9, 42]:

$$T_{cr}(u_m, h) \approx T_C^f \left(1 + \frac{2Q_{12} u_m}{\alpha_T T_C (s_{11} + s_{12})}\right) - \frac{l_S}{\alpha_T \varepsilon_0 \varepsilon_b^f (h + l_S)} \qquad (2.5b)$$

Here, $T_C^f$ is the Curie temperature of bulk ferroelectric, $Q_{12}$ is the component of the electrostriction tensor, $s_{ij}$ are elastic compliances. The positive coefficient $\alpha_T$ is proportional to the inverse Curie-Weiss constant. Since the inequality $s_{11} + s_{12} > 0$ is valid for all ferroelectrics, the positive term $Q_{12} u_m$ increases $T_{cr}(u_m, h)$ and the negative term $Q_{12} u_m$ decreases it. The change of $T_C^f$ can reach several hundred Kelvins [48]. The graphene screening length $l_S$ is usually smaller (or significantly smaller) than 0.1 nm [50, 51]. $\varepsilon_b^f$ is an effective dielectric permittivity of interfacial or "passive" layer on ferroelectric surface [14]. Equation (2.5b) is valid under the condition of an ideal electric contact between the ferroelectric film and graphene channel, and the absence of dead layer is assumed.

**Figure 4(a)** illustrates the situation for a thin BaTiO$_3$ film on tensile substrate with misfit strains $u_m \approx 0.7\%$. $P_S$ increases with thickness and saturates for thin ferroelectric films in the absence of self-polarizing substrate or defects.

Controlling of the Fermi level by temperature changes of the ferroelectric polarization can convert a spin filter to a spin valve. Since $E_F \sim \sqrt{P_S}$ and the temperature dependence of polarization is given by Eqs.(2.5), one yields [42]:

$$E_F(T, u_m, h) = E_F^0 \left(1 - \frac{T}{T_{cr}(u_m, h)}\right)^{1/4}. \qquad (2.6)$$

Here the factor $E_F^0 = \eta v_F \sqrt{\pi P_0 / e}$ can vary in the range $(0.3 - 3.0)$ eV for different ferroelectric films with $P_0 \cong (0.01 - 1)$ C/m$^2$ at temperatures lower (or significantly lower) that $T_{cr}$, the value of Fermi energy in Eq.(6) decreases from $E_F^0$ to 0 with the temperature increase from 0 to $T_{cr}$.

The influence of temperature and film thickness on Fermi energy is presented in **Fig. 4(b)**. It is seen that $E_F(T, u_m, h)$ varies in the range $(0.03 - 0.6)$ eV. Notably, that $T_{cr}(u_m, h)$, should be significantly smaller than the EuO's Curie temperature $T_C$=77 K.

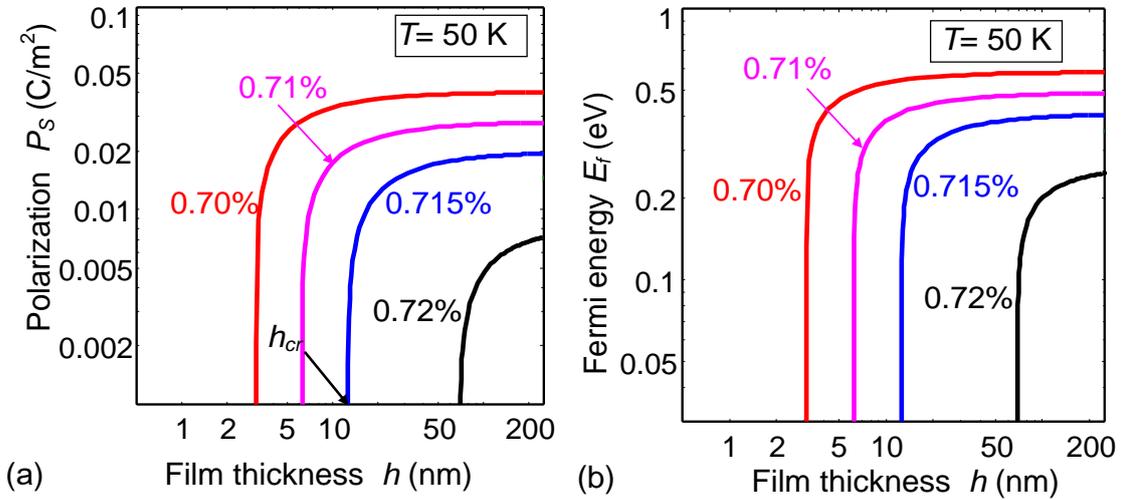

**Fig. 4.** The dependences of the spontaneous polarization $P_S$ of a thin BaTiO$_3$ film **(a)** and the Fermi energy $E_F(T, u_m, h)$ **(b)** on the film thickness $h$ calculated for $u_m \approx 0.7\%$ (red curve), 0.71% (magenta curve), 0.715% (blue curve), and 0.72% (black curve). (Reproduced from [A. I. Kurchak, A. N. Morozovska, and M. V. Strikha. Magnetic dielectric- graphene- ferroelectric system as a promising non-volatile device for modern spintronics J. Appl. Phys. 125, 174105 (2019)], with the permission of AIP Publishing).

In **Fig. 5**, the value of conductivity (2.2) is shown as a function of spontaneous polarization $P_S$ for different values of graphene channel width $W$ = 50, 100 nm and 200 nm. The difference in the form of the curves in **Fig.5** calculated for different values of $W$ is due to the increase in the number of conductance modes occurring with increasing of $W$. Allowing for the assumption about the ballistic nature of transport in the graphene channel, the conductivity depends on the square root of the charge carriers concentration in the channel [16], and, therefore, on the square root of the polarization. If, for sufficiently long channels, the conduction regime becomes diffusive, it will result to the additional factor $\lambda(E_F)/L$ in expression (2), where $\lambda(E_F)$ is the electron free path corresponding to the Fermi level energy [47]. If the scattering occurs predominantly on ionized impurities in substrate (the most common case) then the conductivity will depend linearly on the carrier concentration and polarization [47].

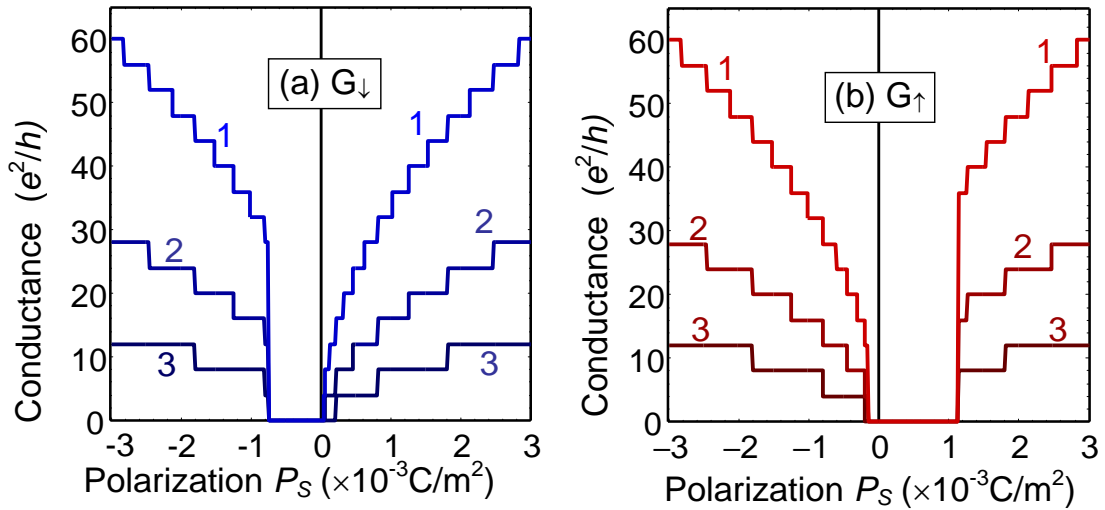

**Fig. 5.** The dependence of the graphene channel conductivity $G_\downarrow$ (**a**) and $G_\uparrow$ (**b**) on the spontaneous polarization $P_S$ calculated for several widths of the graphene channel $W$ = 50, 100 and 200 nm (step-like curves 1, 2 and 3). The conductance is normalized on Klitzing constant $e^2/2\pi\eta$. (Reproduced from [A. I. Kurchak, A. N. Morozovska, and M. V. Strikha. Magnetic dielectric- graphene- ferroelectric system as a promising non-volatile device for modern spintronics J. Appl. Phys. 125, 174105 (2019)], with the permission of AIP Publishing).

## 4. Conclusion

To resume, in Ref.[9] we revealed the unusual features of the IQHE in graphene, unknown earlier. Despite we have an integer character of QHE at both sides of FDW, the total number of perpendicular modes proves to be non-integer in a very unusual way. The nature of this effect principally differs from the traditional non-integer QHE (see e.g. [52]); and the electron gas does not condensate into a special liquid state occurs in this case. Similar non-integer effect has been observed earlier for the p-n junction in graphene channel created by two gates [12], however the presence of ferroelectric substrate with FDWs modifies the character of non-integer QHE and introduces new smart details into it.

In [10] we considered the conductivity of the magnetic dielectric placed on the graphene conductive channel, which in turn was deposited at the ferroelectric substrate. The magnetic dielectric locally transforms the band spectrum of graphene by inducing an energy gap in it and making it asymmetric with respect to the spin of the free electrons. The range of spontaneous polarization of ferroelectrics (2 – 5) mC/m², which can be easily realized in thin (10 – 100) nm films of proper and incipient ferroelectrics, was under examination. If the Fermi level in the graphene channel belongs to energy intervals where the graphene band spectrum, modified by EuO, becomes sharply spin-asymmetric, such a device can be an ideal non-volatile spin filter. However, it cannot operate without the top gate.

Note, that the problem solved in Ref.[10] has a framework character. The practical application of the system under consideration is restricted by a relatively low ferromagnetic transition temperature of EuO, $T_C^m = 77 K$. However, as it was demonstrated by Hallal et al. [53], alternative magnetic insulators with higher Curie temperatures can cause similar local transformation of graphene band spectrum. According to the first principles calculations [53], energy gaps, imposed in graphene by magnetic insulator $Y_3Fe_5O_{12}$ (YIC), are similar to the ones described by Eq.(2.2); however, the "useful" energy ranges with spin asymmetry are several times wider there, which makes such a system more convenient for practical usage. High ferromagnetic transition temperature of YIC ($T_C^m = 550 K$) permits the system to operate under ambient conditions. However, quite recently Song [54] demonstrated that the gate-induced spin valve based on graphene/YIG (or on graphene/EuS) also induces a heavy electron doping, 0.78eV, which corresponds to a giant spontaneous polarization of 1.5 C/m². Therefore, a system magnetic insulator – grapheme – ferroelectric can be treated as a promising one for spintronic devices of new generation.

**Acknowledgments.** A.N.M. work was partially supported the European Union's Horizon 2020 research and innovation program under the Marie Skłodowska-Curie (grant agreement No 778070).